\documentclass[aps,prl,reprint,superscriptaddress,floatfix,showpacs]{revtex4-1}
\usepackage{amssymb,bm}
\usepackage{amsmath,eucal,ulem}
\usepackage[usenames,dvipsnames]{color}
\usepackage{graphicx}

\begin{document}

\title{Entropic anomaly and maximal efficiency of microscopic heat engines} 
\author{Stefano Bo}
\affiliation{Cancer Cell Biophysics, IRC@C: Institute for Cancer Research at Candiolo, Str.~Prov.~142~km~3.95, 10060 Candiolo,
Torino, Italy}
\affiliation{INFN, via P. Giuria 1, 10125 Torino, Italy} 
\author{Antonio Celani}
\affiliation{Physics of Biological Systems, Institut Pasteur and CNRS UMR 3525, 28 rue du docteur Roux, 75015 Paris, France} 
\begin{abstract}
The efficiency of microscopic heat engines in a thermally heterogenous environment 
is considered. We show that -- as a consequence of the recently discovered entropic anomaly -- quasi-static engines, whose efficiency is maximal in a fluid at uniform temperature,
have in fact vanishing efficiency in  presence of temperature gradients. For slow cycles the efficiency falls off as the inverse of the period. 
The maximum efficiency is reached at a finite value of the cycle period that is
 inversely proportional to the square root of the gradient intensity. The relative loss in maximal efficiency with respect to the thermally
 homogeneous case grows as the square root of the gradient. As an illustration of these general results, we construct an explicit, analytically solvable example of a 
Carnot stochastic engine. In this thought experiment, a Brownian particle is confined by a harmonic trap and immersed in a fluid with a linear temperature profile. This example may serve as a template for the design of real experiments in which  the effect of the entropic anomaly  can be measured.

\end{abstract}
\date{\today}
\pacs{05.40.-a, 05.70.Ln}

\maketitle

\paragraph{Introduction.---}
Motion at the microscopic scales is dominated by viscous resistance \cite{P77}. 
In this Aristotelic world, inertia appears to play a diminutive role. As such, it seems justified to
disregard it altogether. Such {\it overdamped approximation} has been the cornerstone
of the theory of Brownian motion since Einstein, Smoluchowski and Langevin. 
It therefore comes as a surprise the recent finding that
while microscale mechanics conforms well to the overdamped approximation thermodynamics                                                 
does not, when a thermally heterogeneous environment is considered \cite{CBEA12}. 
In this case, thermodynamical quantities evaluated in the limit of arbitrarily small yet finite inertia do not coincide with their counterparts for exactly zero inertia. Most notably,
the limiting average entropy production for a Brownian particle is
\begin{equation}\label{eq:anom}
\lim_{m\to 0} \frac{d\left\langle S_{tot} \right\rangle}{dt}  =
\frac{d\left\langle S_{tot} \right\rangle}{dt} \biggr|_{0} +  \frac{(n+2)k_B^2}{6} 
\left\langle \frac{(\nabla T)^2}{\gamma T}  \right\rangle_{0}
\end{equation}
where $m$ is the particle mass, the subscript $_0$ denotes quantities evaluated in the 
overdamped approximation ($m=0$), $T$ is the temperature, $n$ is the spatial dimension, $\gamma$ is the friction 
coefficient, and $k_B$ is the Boltzmann constant \cite{CBEA12,Note0}.

This singular limit has been dubbed {\it entropic anomaly}. Its 
theoretical origin has been traced back to the breaking of a fundamental symmetry of the overdamped dynamics when moving to the small, yet finite, inertia limit 
\cite{CBEA12}. The additional positive contribution to entropy production 
signals that essentially irreversible processes are at work
at the microscopic scale in a thermally inhomogeneous environment. 
Indeed, on fast timescales of the order of $m/\gamma$,  the particle sways back and forth
over a lengthscale $(k_B T m)^{1/2}/\gamma$. During these oscillations, heat is 
more likely absorbed when the particle is at a higher temperature and released at a smaller one, thereby producing entropy. Even if the amount of entropy produced in each of these futile cycles is small,  $\sim m$, their frequency is so high, $\sim m^{-1}$,
that it leads to a finite entropy production even in the limit of vanishing mass \cite{CBEA12}. 

Therefore, in view of the tight link between thermodynamic efficiency  of heat engines and entropy production which follows from the second law of thermodynamics, we asked ourselves if and how the entropic anomaly has an impact on the efficiency of stochastic heat engines. 

In this Letter we show that anomalous entropy production dramatically affects 
the efficiency of microscopic heat engines. The most prominent effect is seen 
in the quasi-equilibrium 
regime, where efficiency should be maximal -- and actually is in the absence of temperature gradients.
The anomaly induces a crisis and a fall-off of efficiency proportional to the inverse
of the cycle period. The maximal efficiency is therefore reached at a finite cycling time.
We find that the relative loss in maximal efficiency is proportional to the square root of 
the gradient. The maximum occurs at a cycle duration that is proportional to the inverse square root of the gradient. These results are exemplified by the analytical solution of a stochastic Carnot engine in a linear gradient. Motivated by recent experiments that have realized similar microscopic engines by means 
 of optical tweezers and laser heating (see e.g.~\cite{BB11}) we show that the efficiency fall-off due to the anomaly may be accessed by experimental measurements. 
In the main text below we report the most important results, leaving the derivations for the Supplementary Information.

\paragraph{ Mechanics and thermodynamics at the microscale.---}
The motion of a microscopic particle in a fluid is described by the Langevin-Kramers equations 
\begin{equation}\label{eq:LK}
\begin{array}{lll}
\dot{X}^i_t &=& V^i_t \\
m \dot{V}^i_t &=& f^i -\gamma V_t^i  + 
\sqrt{2 k_B T \gamma}\, \eta^i_t 
\end{array}
\end{equation}
where $X^i_t$ and $V^i_t$ are position and velocity coordinates in $n$-dimensional space
and $\eta^i_t$ are independent white noises.  In the following we shall consider a 
generic dependence of force $f^i(X_t,t)$, temperature $T(X_t,t)$ and friction coefficient $\gamma(X_t,t)$ on space and time, unless explicitly stated otherwise. Here and throughout the text, the products are always taken with the Stratonovich midpoint convention. 

We now introduce some basic notions needed to describe the thermodynamics of stochastic heat engines \cite{Sekimoto,Seifert_review}.
For a conservative force $f_i=-\partial U/\partial x^i$ the power exerted on the particle along the trajectory is $\dot{W} = 
\partial U /\partial t$ and the rate of heat release is $\dot{Q}=f^i \dot{X}^i - m V^i \dot{V}^i $, so that 
the first law of thermodynamics reads $d(U + mV^iV^i/2)/dt = \dot{W}-\dot{Q}$  .
Upon defining the  state entropy
$S_p=-\log p $,  where $p$ is the solution of the Fokker-Planck equation associated to \eqref{eq:LK} (see Ref.~\cite{Seifert,CG08}), and the rate of entropy production in the fluid environment as $\dot{S}_{env} = \dot{Q}/T$, the second law of stochastic thermodynamics states that the average rate of entropy production  $d\langle S_{tot}\rangle/dt=d\langle S_p \rangle/dt  + d\langle S_{env} \rangle/dt$ always increases, vanishing only at equilibrium. 


\paragraph{Entropy production and efficiency.---}
Irreversibility, in the form of positive total entropy production, sets a fundamental limit to the efficiency of stochastic cyclic heat engines, just as it does for ordinary macroscopic thermodynamics. Efficiency is defined as the ratio of extracted work to absorbed heat
during a cycle
\begin{equation}\label{eq:eff}
\eta = \frac{\langle W_{extr}\rangle}{\langle Q_{abs}\rangle} = 1-\frac{\langle Q_{rel}\rangle}{\langle Q_{abs}\rangle}
\end{equation}
Explicitly, introducing the probability $\hat{p}=\langle \delta(T-\hat{T})\rangle$ that the particle is in contact with the thermostat at temperature $\hat{T}$, and the heat release rate to
the same thermostat $\dot{q} = \hat{p}^{-1} \langle \dot{Q} \delta(T-\hat{T})\rangle$, 
one has
$\langle Q_{rel}\rangle=\oint dt \int d\hat{T}\,\hat{p}\dot{q} \theta(\dot{q}) $,
$\langle Q_{abs}\rangle=-\oint dt \int d\hat{T}\,\hat{p}\dot{q} \theta(-\dot{q})$
and $\langle W_{extr}\rangle=\langle Q_{abs}\rangle-\langle Q_{rel}\rangle$ by the first law.
The total entropy production during a cycle then reads $\langle S_{tot} \rangle = \langle S_{env} \rangle= \oint dt \int d\hat{T}\, \hat{p}\dot{q}/\hat{T}$. When the temperature is uniform, the usual definitions of all these quantities are recovered.
Efficiency can be rewritten in a highly suggestive form that exposes its relationship
with irreversibility
\begin{equation}\label{eq:eff-entr}
\displaystyle
\eta = \frac{1-\frac{T_{rel}}{T_{abs}}} {1+\frac{T_{rel} \langle S_{tot}\rangle  }{\langle W_{extr}\rangle} }
\end{equation}
where  
$ T_{rel/abs} = Q_{rel/abs}/S_{rel/abs} $ 
are cycle-averaged temperatures weighted by the entropy release and absorption rates, $S_{rel} =
\oint dt \int d\hat{T}\, \hat{p}\dot{q} \theta(\dot{q})\hat{T}^{-1} d\hat{T}$ and 
$S_{abs} =-
\oint dt \int d\hat{T}\,\hat{p}\dot{q} \theta(-\dot{q})\hat{T}^{-1} d\hat{T}$.

For a stochastic heat engine that operates in a spatially {\it homogeneous} temperature environment, between
temperatures $T_c$ and $T_h$, the maximal efficiency is the Carnot efficiency
$\eta_c =1 -T_c/T_h$. It follows from Eq.~\eqref{eq:eff-entr} that it can be reached only if: {\it (i)} the engine works
reversibly ($S_{env}=0$); {\it (ii)} heat is absorbed at the higher temperature $T_h$ only; 
 {\it (iii)} heat is released at the lower temperature $T_c$ only. The requirement of reversibility 
 implies that the transformations must be performed at quasi-equilibrium, therefore 
in the limit of infinitely slow cycling and in the absence of fluxes. In this regime,
the other two conditions can be met by a sequence of isothermal and adiabatic transformations
akin to the classical Carnot cycle. An explicit implementation of a stochastic Carnot engine is given below.

\paragraph{The efficiency crisis.---}
When the environment is thermally {\it inhomogeneous}, it is not possible to reduce the 
entropy production to nil, because of the entropic anomaly in Eq.~\eqref{eq:anom}. As a result, Carnot efficiency
cannot ever be reached. Additionally, since the contribution to entropy production  from the anomalous term is proportional to the cycle duration, whereas for large periods all other terms in Eq.~\eqref{eq:eff-entr} become independent of it, we expect that the efficiency 
will eventually fall off as the inverse of the cycle period. This decay occurs even for
transformations that are conducted at overdamped quasi-equilibrium dynamics, i.e. for $\langle S_{tot}\rangle_{0}=0$. Therefore, the asymptotic efficiency tends to zero as the inverse of the cycle duration $\tau$
\begin{equation}\label{eq:effanom}
\eta \simeq \frac{ W_{extr}^{\infty} }{ T_{rel}^{\infty} \langle S_{anom}\rangle}
\left( 1-\frac{T_{rel}^{\infty}}{T_{abs}^{\infty}} \right) 
\simeq \frac{\tilde\tau}{\tau}  \qquad \mbox{for  } \tau \gtrsim \tilde\tau
\end{equation}
where 
$\langle S_{anom}\rangle=(n+2)k_B^2 \oint dt\langle (\nabla T)^2/(6\gamma T)\rangle$.
This reduction in efficiency is due to the microscopic futile cycles discussed in the introduction. These are responsible for anomalous entropy production and constitute the major contribution to heat absorption at large cycling times. 

For shorter cycles, the anomalous entropy contribution is negligible with respect to 
the overdamped one, and the efficiency tends to attain its value in absence of
temperature inhomogeneities
\begin{equation}\label{eq:effg0}
\eta \simeq \eta_\infty \left(1-\frac{\tau_r}{\tau}\right) \qquad \mbox{for  }
\tau_r \ll \tau \lesssim \tau_*
\end{equation}
where $\tau_r$ is the characteristic relaxation time to equilibrium and $\eta_\infty$ 
is the maximal efficiency in the quasi-equilibrium, homogeneous case (e.g. $\eta_c$ 
for a Carnot engine). The previous expression is just the first order approximation to
efficiency upon developing in the small parameter $\tau_r/\tau$.

In summary, we expect efficiency to grow for increasing cycle durations until a first crossover time $\tau_*$ and then to decrease for durations longer than $\tilde{\tau}$, reaching a maximum value 
at a finite period $\tau_{max}$ comprised between the two.   
\begin{figure}[!t]
\includegraphics[angle=90,width=\columnwidth,trim=100 100 100 100]{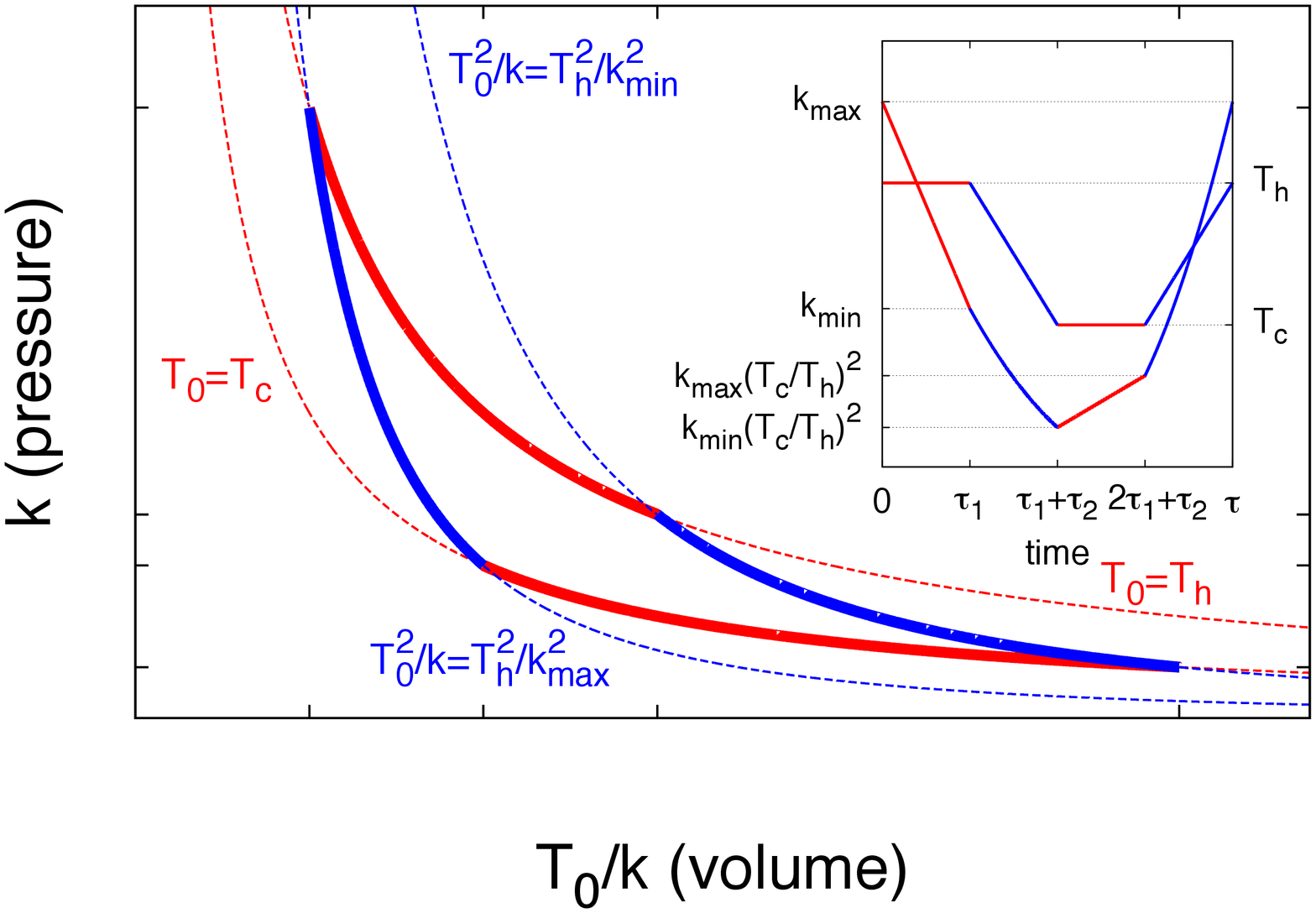}
\caption{The stochastic Carnot engine.
We adopted the following protocol: 
{\it (i)} isothermal expansion of duration $\tau_1$ at $T_h$ with linearly decreasing stiffness from $k_{max}$ to $k_{min}$;  
{\it (ii)} adiabatic expansion of duration $\tau_2$ with linear decrease in temperature from $T_h$ to $T_c$ and $k=k_{min}(T_0/T_h)^2$; 
{\it (iii)} isothermal compression of duration $\tau_1$ at $T_c$ 
with linearly increasing stiffness from $k_{min}(T_c/T_h)^2$ to $k_{max}(T_c/T_h)^2$;
{\it (iv)} adiabatic compression of duration $\tau_2$ with linear increase in temperature from $T_c$ to $T_h$ and $k=k_{max}(T_0/T_h)^2$. The overall cycle duration is $\tau=2\tau_1+2\tau_2$.}
\label{fig:1}
\end{figure}

\paragraph{Maximal efficiency of a stochastic Carnot engine in a thermally inhomogeneous 
environment.---} 
In this section we give theoretical estimates for the the maximum efficiency $\eta_{max}$, and corresponding duration $\tau_{max}$, for a stochastic Carnot engine in
the physically relevant case of temperature gradients $g$ of small intensity $g L/T \ll 1$
where $L$ is a typical value of standard deviation of particle position.
The onset of the efficiency crisis occurs during adiabatic transformations. There, the amount
of heat absorbed -- because of irreversibility -- falls off as the inverse duration of the transformation $\sim \tau_r/\tau$. This decrease persists until the anomalous absorbed heat,
which increases proportionally to the duration $\sim \tau/\tilde{\tau}$, takes over.   
The corresponding increase in heat absorption determines a small, yet steady, linear 
decrease in efficiency for larger cycle durations. It is therefore at this crossover that
maximal efficiency is reached
\begin{equation}\label{eq:taumax}
\tau_{max} \sim \tau_* \sim \left(\tau_r \tilde{\tau}\right)^{1/2} \ll \tilde\tau
\end{equation}
At even larger durations $\tau \gtrsim \tilde\tau$, 
the anomalous term eventually overcomes the contribution from the isothermal expansion as well, and efficiency falls off as the inverse of cycle duration.
Finally, for small gradients and large cycle periods, anomalous heat release and absorption
occur at any time in the cycle at temperatures $\sim T_0 \mp g L$.  The anomalous entropy production is $S_{anom} \sim \tau k_B^2 g^2/(\gamma \bar{T}) $ whence, 
using Eqs.~\eqref{eq:effanom} and \eqref{eq:taumax} one obtains the estimates
\begin{equation}\label{eq:tildetau}
\tilde\tau \sim \frac{\gamma L}{k_B g} \qquad \tau_{max} \sim 
\left( \frac{\gamma \tau_r L}{k_B g} \right)^{1/2} 
\end{equation}
The maximal efficiency follows from combining Eqs.~\eqref{eq:effg0}, \eqref{eq:taumax}
and \eqref{eq:tildetau}
\begin{equation}\label{eq:etamax}
\frac{\eta_c -\eta_{max}}{\eta_c} \sim \frac{\tau_r}{\tau_{max}} \sim \left( \frac{k_B g \tau_r} {\gamma L}\right)^{1/2}
\end{equation}

\begin{figure}[!ht]
\centering
\includegraphics[width=\columnwidth,trim=50 50 50 50]{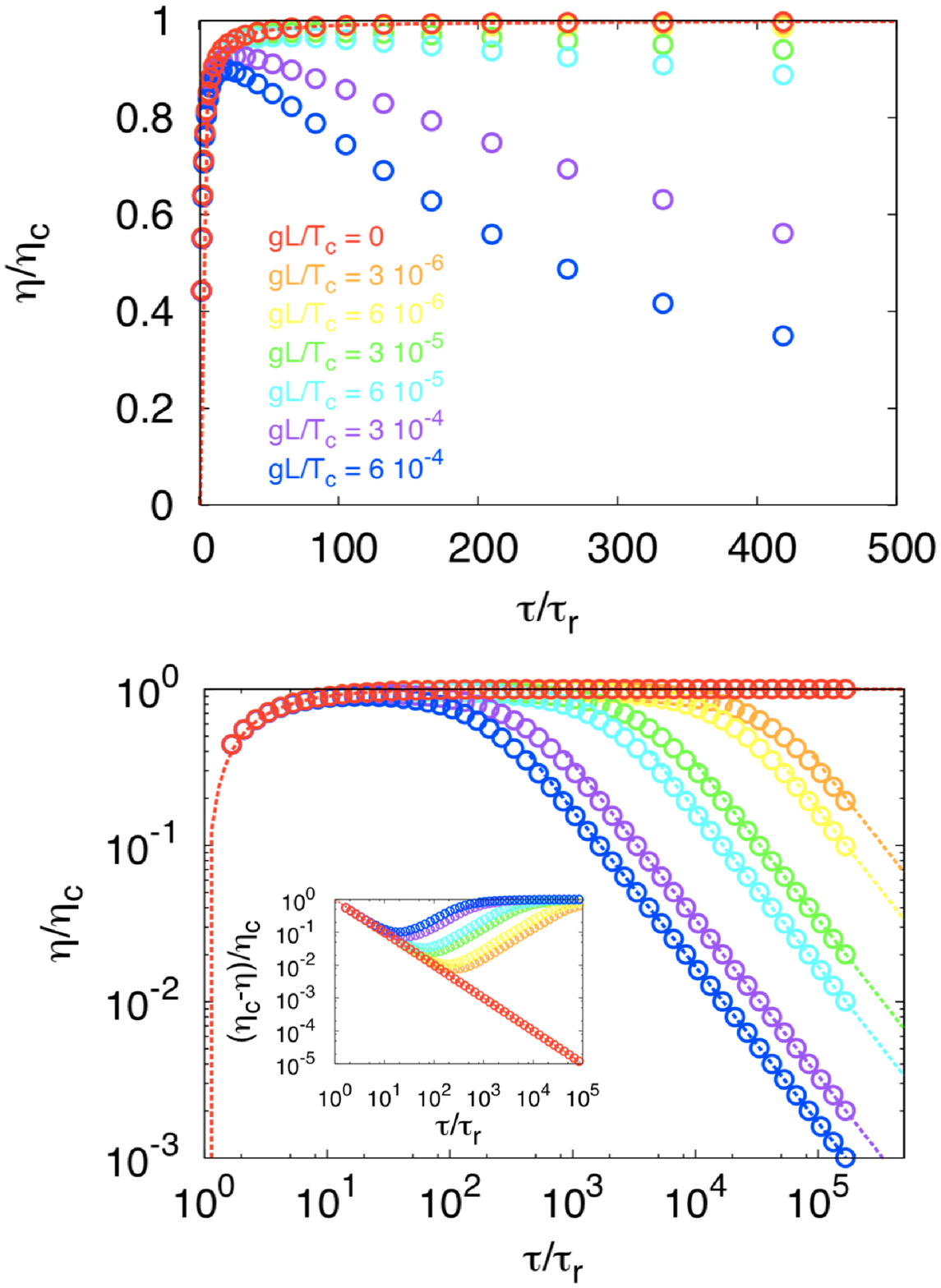}
\caption{
The efficiency crisis. Top: In the absence of gradients, the efficiency reaches the maximal Carnot value in the quasi-equilibrium (red circles). When a gradient is present, the efficiency falls short of approaching $\eta_c$ and eventually decreases to zero for large cycling times.
Bottom: Same data as in the upper panel, now on a doubly logarithmic scale. The dashed lines
are given by Eq.~\eqref{eq:effanom}. The exact analytic expression for $\tilde{\tau}$, which conforms to the scaling in  Eq.~\eqref{eq:taumax}, is given in the Supplementary Information. The red dashed line is Eq.~\eqref{eq:effg0}, which, as shown in the inset, provides a very good approximation of efficiency for cycling times $\tau_r \lesssim \tau \lesssim \tau_*$.}
\label{fig:2}
\end{figure}


\begin{figure}[!h]
\centering
\includegraphics[width=\columnwidth,trim=20 20 20 20]{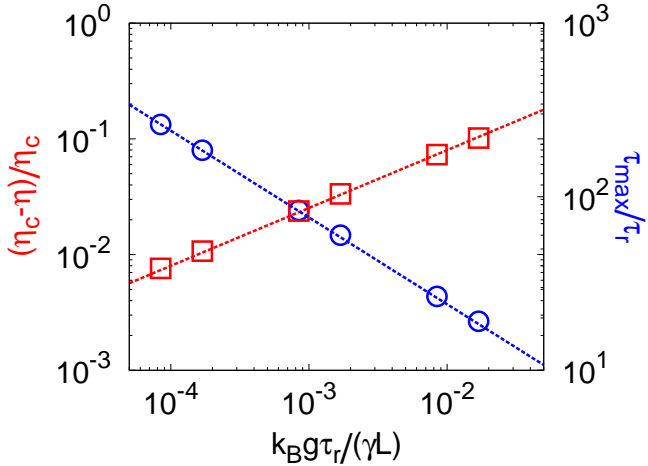}
\caption{
The maximal efficiency of the stochastic Carnot engine as a function of the gradient intensity. Dashed lines are the square-root and inverse-square-root predictions from Eqs.~\eqref{eq:etamax} and \eqref{eq:taumax}. 
}
\label{fig:3}
\end{figure}

\paragraph{A stochastic Carnot engine: harmonic trap in a linear temperature profile.---}
As an illustrative example of the general arguments given above, here we offer a detailed
analysis of a stochastic heat engine that is able to attain Carnot efficiency in a a thermally homogeneous environment and
displays the efficiency crisis when subject to a temperature gradient. We consider a Brownian
particle, confined by a harmonic trap of stiffness $k$ and immersed in a fluid with a linear temperature profile $T=T_0+gx$. For the sake of simplicity we consider a space-independent friction coefficient $\gamma$ and we limit ourselves to the one-dimensional case. For small gradient amplitudes $g\sigma/T_0 \ll 1$, where $\sigma^2=\langle x^2\rangle$ is the variance in particle position, it is possible to obtain analytical expressions at all times
for the probability of being in contact with the thermostat $\hat{T}$  
\begin{equation}\label{eq:pharm}
\hat{p}
= \frac{e^{-\frac{(\hat{T}-T_0)^2}{2g^2\sigma^2} }}{(2\pi g^2\sigma^2)^{1/2}}  \left[ 1 + \frac{g\sigma}{3 T_0} \left(\frac{(\hat{T}-T_0)^3}{g^3\sigma^3}-3\frac{\hat{T}-T_0}{g\sigma} 
\right) \right]   + \ldots
\end{equation} 
and the average heat release rate to that thermostat 
\begin{equation}\label{eq:qdotharm}
\begin{array}{lll}
\dot{q} &=& -k_B T_0 \frac{\dot{\sigma}}{\sigma} -\frac{k_B}{2} \dot{T}_0 + 
\gamma\dot{\sigma}^2 \left(\frac{\hat{T}-T_0}{g\sigma}\right)^2 \\ & & 
- \left(\frac{g\sigma}{T_0} \right) \left( 2 k_B \frac{\dot\sigma T_0}{\sigma} + \frac{k_B^2}{2} \frac{T_0^2}{\gamma\sigma^2}
\right) \left(\frac{\hat{T}-T_0}{g\sigma}\right)  + \ldots
\end{array}
\end{equation}
Since the standard deviation of position evolves according to $\gamma\dot{\sigma}=-k\sigma + T_0/\sigma$,
it is then possible to obtain by quadratures the total absorbed heat during a cycle, 
the extracted work 
, and the ensuing efficiency. As for the definition of the Carnot engine protocol, a crucial ingredient is the 
definition of an adiabatic transformation. From the inspection of  
Eq.~\eqref{eq:qdotharm} it appears that in the absence of gradient and in the quasi-static limit
(i.e. $\gamma|\dot\sigma|/(k \sigma) \ll 1$) the heat release rate reduces to $\dot{q}_\infty
= -k_B [k/(2T_0)] \frac{d}{dt}(T_0^2/k)$. Therefore, quasi-static transformations that keep the ratio $T_0^2/k$ constant are adiabatic. Notice that this definition correctly accounts for
the kinetic energy contributions \cite{Seifert2}. 
Accordingly, we construct a Carnot
engine by a sequence of two isothermal and two adiabatic transformations (see Figure~\ref{fig:1}). 

As
shown in Figure~$\ref{fig:2}$, in the absence of temperature gradient the efficiency hyperbolically approaches Carnot efficiency. Conversely, in the presence of gradient, the efficiency undergoes
a crisis at cycle periods $\sim \tau_*$ and eventually rapidly falls-off for durations $\gtrsim \tilde\tau$, as predicted in the previous section.  
It is worth noting that in this last regime, 
Eq.~\eqref{eq:qdotharm} reduces to $\dot{q} \simeq \dot{q}_{anom} =-  k(2\gamma)^{-1} [ k_B (\hat{T}-T_0) ]$ and heat is absorbed on the side of the trap at higher temperature and released at the opposite one, in equal amounts, at all times during the cycle.
In Figure~\ref{fig:3} it is shown the dependence of the maximal efficiency, and of the cycle period at which it occurs, on 
the gradient strength. Both agree very well with Eqs.~\eqref{eq:taumax}
and \eqref{eq:etamax}.
\paragraph{Conclusions and discussion.---}
We have shown how the entropic anomaly of stochastic thermodynamics discovered in Ref.~\cite{CBEA12} affects the efficiency of microscopic engines. To this aim we have designed a {\it gedanken experiment} where a particle is trapped in a harmonic potential and subject to a linear temperature gradient. For such a system the efficiency of a stochastic engine can be computed analytically.
By a suitable definition of adiabatic transformations, we have shown that it is possible to construct a Carnot engine, at least in absence of gradients. When gradients are present,
the efficiency undergoes a crisis and decays to zero for large cycling periods because 
of anomalous, irreversible heat absorption. It is then natural to ask whether the anomaly effect may or may not have a tangible impact on real experiments. When 
actual experimental parameters are considered,
one has $T\simeq 300 \, ^{\circ}$K,
$L \simeq 0.2 \,\mu$m and $g\simeq 1\, ^{\circ}$K/$\mu$m \cite{BB11,C09}, and $gL/T\sim 6\cdot 10^{-4}$. Under these conditions, the Carnot engine reaches a maximal efficiency for periods $\sim 150$ seconds, at
about $90 \%$ of the Carnot value, and then falls off because of the anomalous heat 
absorption. This estimate suggests that an experimental verification of the entropic anomaly is within reach of current experimental capabilities, prompting an effort in this direction.

\end{document}